\documentstyle[11pt,newpasp,twoside,epsf]{article} 
\def\edcomment#1{\iffalse\marginpar{\raggedright\sl#1\/}\else\relax\fi}
\reversemarginpar

\begin{document}
\title{The autocorrelation function of the soft X-ray background}
\author{Andrzej M.\ So\l tan}
\affil{Copernicus Astronomical Center, Bartycka 18, 00-716 Warsaw, Poland}
\author{Michael J.\ Freyberg}
\affil{MPI f\"ur extraterrestrische Physik, D-85748 Garching, Germany}

\begin{abstract}
The first positive detection of the X-ray background fluctuations at small
angular scales is reported. {\it ROSAT} PSPC archive pointed observations
are used to measure fluctuations at scales of $0\fdg03 - 0\fdg4$. The pointings
have been selected from an area free from galactic contamination. At
separations below $\sim 0\fdg1$ clusters of galaxies become a substantial
source of the background fluctuations. The autocorrelation function of
the fluctuations in the power law approximation has a slope of $\sim 1$
for all the data but is substantially flatter (with slope of $\sim 0.7$)
when pointings containing bright clusters are removed. At separations
$0\fdg3-0\fdg4$ where the ACF estimates based on the ROSAT pointings and
All-Sky Survey are available, both data sets give consistent results.
\end{abstract}

\section{Introduction}

Fluctuations of the X-ray background (XRB) provide a unique method
to investigate statistical properties of the distribution of X-ray
sources. At angular scales above $\sim 0\fdg3$ the {\it ROSAT} All-Sky
Survey (RASS) revealed distinct variations of the soft XRB (Soltan et
al.\ 1996). Subsequent analysis (Soltan et al.\ 1999) showed that clustering
of active galactic nuclei (which are the main contributor to the XRB)
potentially could explain the large amplitude of the XRB autocorrelation
function (ACF). However, the amplitude of X-ray source spatial clustering
required to reproduce the observed ACF is higher than that derived from
direct investigation of the distribution of sources (e.g.\ Carrera et
al.\ 1998). Thus, it is likely that fluctuations of the soft XRB are
produced also by some other effects. In particular, hydrodynamical
computations by Cen \& Ostriker (1999) show that the process of accumulation
of primordial gas in the potential wells created by galaxies and clusters
of galaxies produced large scale concentrations of hot plasma.
This material is expected to emit thermal bremsstrahlung in the soft
X-ray domain. A clumpy distribution of emitting plasma would contribute
substantially to the total fluctuations of the XRB. To investigate the
importance of different mechanisms producing the XRB anisotropy 
one needs to measure the amplitude of the XRB fluctuations over a wide
range of angular separations and at different energies. In the present
work we have determined the autocorrelation function of the soft
XRB using a large number of {\it ROSAT} pointings. Below we describe the
observational material used in the analysis, the computational details
and compare the present results with our earlier measurements of the
ACF at large separations using the {\it ROSAT} All-Sky Survey.

\section{Analysis of the observational data}

Visual inspection of the RASS X-ray maps (SNowden et al.\ 1997) 
shows that at low energies
only selected sky regions seem to be free from thermal emission
by hot galactic plasma and not affected by absorption by cold gas. To
investigate fluctuations of the extragalactic component of the XRB Soltan
et al.\ (1996) used a section of the RASS maps of approximately 1 sr at
the northern galactic hemisphere ($b > 40\deg, 70\deg < l < 250\deg$)
apparently least contaminated by local effects. In the present analysis
we have concentrated on the same region of the sky. All {\it ROSAT}
pointed observations within this area with exposure time longer than
5000\,s excluding pointings at known extended sources (supernova remnants,
nearby normal galaxies, clusters of galaxies) have been used. Total
number of pointings satisfying these selection criteria is equal to
141. The data in the gain-corrected pulse height channels $91-131$
('R6 band' in Snowden at al.\ 1994) with the energy centered at
1.15\,keV has been selected. Counts in each field were binned into
$8\arcsec\times8\arcsec$ pixels. For each observation the central region
containing the target source was eliminated from the data, the size of the
removed area was carefully determined. The distribution of counts within
the field of view produced by the target source was calculated using
the {\it ROSAT} telescope point spread function and the radius of the circle
to be removed was obtained in such a way that the residual contribution
of the target source did not exceed 5\,\% of the local background within
the remaining section of the field of view.

To minimize vignetting effects we have used the central section of the
PSPC field of view with radius of 13\farcm5. Contamination by particle
background and by soft solar photons scattered in the Earth's atmosphere
was subject to scrutiny for each observation separately. In the vast
majority of observations the number of non-cosmic counts was insignificant,
and in few cases the observed count rates have been corrected for both
these effects.

\begin{figure}
\plotfiddle{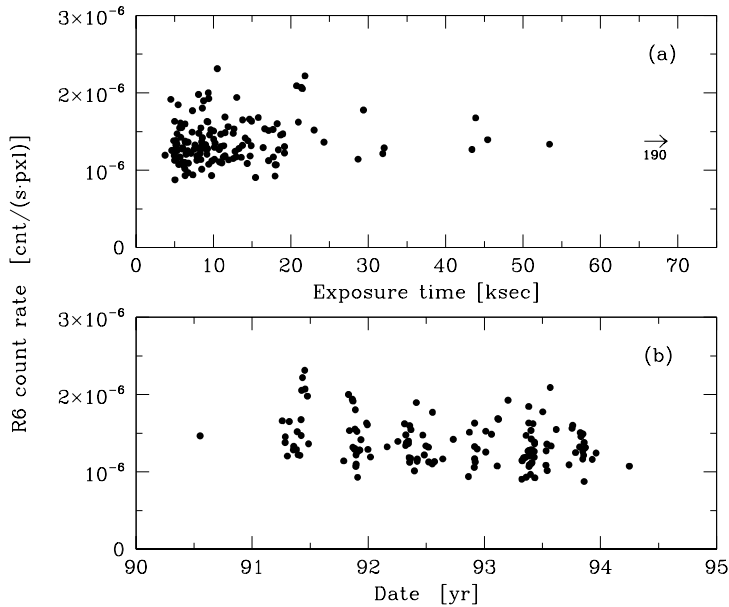}
{70mm}{0}{50}{50}{-150}{-120}
\caption{Distribution of count rates in the sample vs.\ exposure time (a),
and vs.\ observation date (b).}
\end{figure}

It is of fundamental importance for the fluctuation analysis to construct
a sample of observations for which all the instrumental and non-cosmic
effects are constant and do not vary from one observation to another. To
check for the possible systematic variations of the telescope/detector
system characteristics we have plotted in Fig.\,1\,a and b respectively,
the average count rate per pixel as a function of exposure time and date
of observation.  While the count rates averaged over each field do not
show any dependence on exposure time, they vary systematically with
the date of the observation. Most probably this effect results from
small inaccuracies in the gain corrections. To remove this weak trend,
a straight line was fitted to the data and the average count rate in each
field was corrected for the slope of the fit.

We denote the intensity of the XRB in the direction defined by the unit vector
$\bf n$ in the $i$-th pointing (= count rate in the corresponding
pixel) as $\rho_i({\bf n})$. The ACF $w(\theta)$ was calculated using
the formula:
\begin{displaymath}
w(\theta)=N\,{\Sigma_i^N \langle\rho_i({\bf n}) \rho_i({\bf n^\prime})\rangle
                \over \left[ \Sigma_i^N\langle \rho_i \rangle \right]^2} -1\,,
\end{displaymath}
where the sum extends over all pointings ($N=141$), the directions ${\bf n}$
and ${\bf n^\prime}$ are separated by the angle $\theta$, and
$\langle \rho_i\rangle$ denotes the count rate in the $i$-th field
calculated averaged over pixels used in the calculations for given
$\theta$.

\section{Autocorrelation function from pointings and RASS}

The ACF based on the pointings is shown in Fig.\,2a. The strong signal at
separations below $\sim 100\arcsec$ results from ``granular'' nature
of the XRB. The shape of the ACF at these separations is defined by
the PSF of the telescope/detector system averaged over the field of view
and photon energies. Variations of the ACF at larger separations also
result from the discrete structure of the XRB. Here the major role play
pairs (or multiplets) of strong sources in each field; the larger the number
of separate pointings used in the calculations, the smaller amplitude of
the ACF fluctuations.

Apart from the fluctuations, the ACF amplitude shows a systematic decline
with increasing separation. Above $\sim 1000\arcsec$ the
signal becomes too weak in comparison with the fluctuations produced
by the strong sources, that the ACF cannot be estimated reliably.

Present measurements of the ACF are shown together with the RASS
estimates (taken from Soltan et al.\ 1999) in Fig.~2b (note the logarithmic
scale). Open circles correspond to the results shown in Fig.~2a averaged
in logarithmic bins. Both measurements are in satisfactory agreement
taking into account large uncertainties of both measurements (error
bars of the RASS result represent scatter due to statistical fluctuations).

The ACF signal is now measured in the range $0\fdg03 - 10\deg$.
It is likely that various effects contribute the XRB fluctuations
at different scales. To assess what role play the bright clusters of
galaxies we have repeated all the calculations without fields containing
known clusters\footnote{Pointings with known clusters as observational targets
have been removed from the sample in the preliminary selection of the
material; here we consider only serendipitous clusters in the
field of view.}. The ACF calculated for the restricted data set (133 pointings)
is shown in Fig.~2b with crosses. It appears that clusters contribute to the
XRB fluctuations noticeably only at separations below $\sim 0\fdg1$.
The ACF without known clusters becomes substantially flatter than for
all the data. Detailed models of the ACF will be discussed in a separate
paper.

\begin{figure}
\plottwo{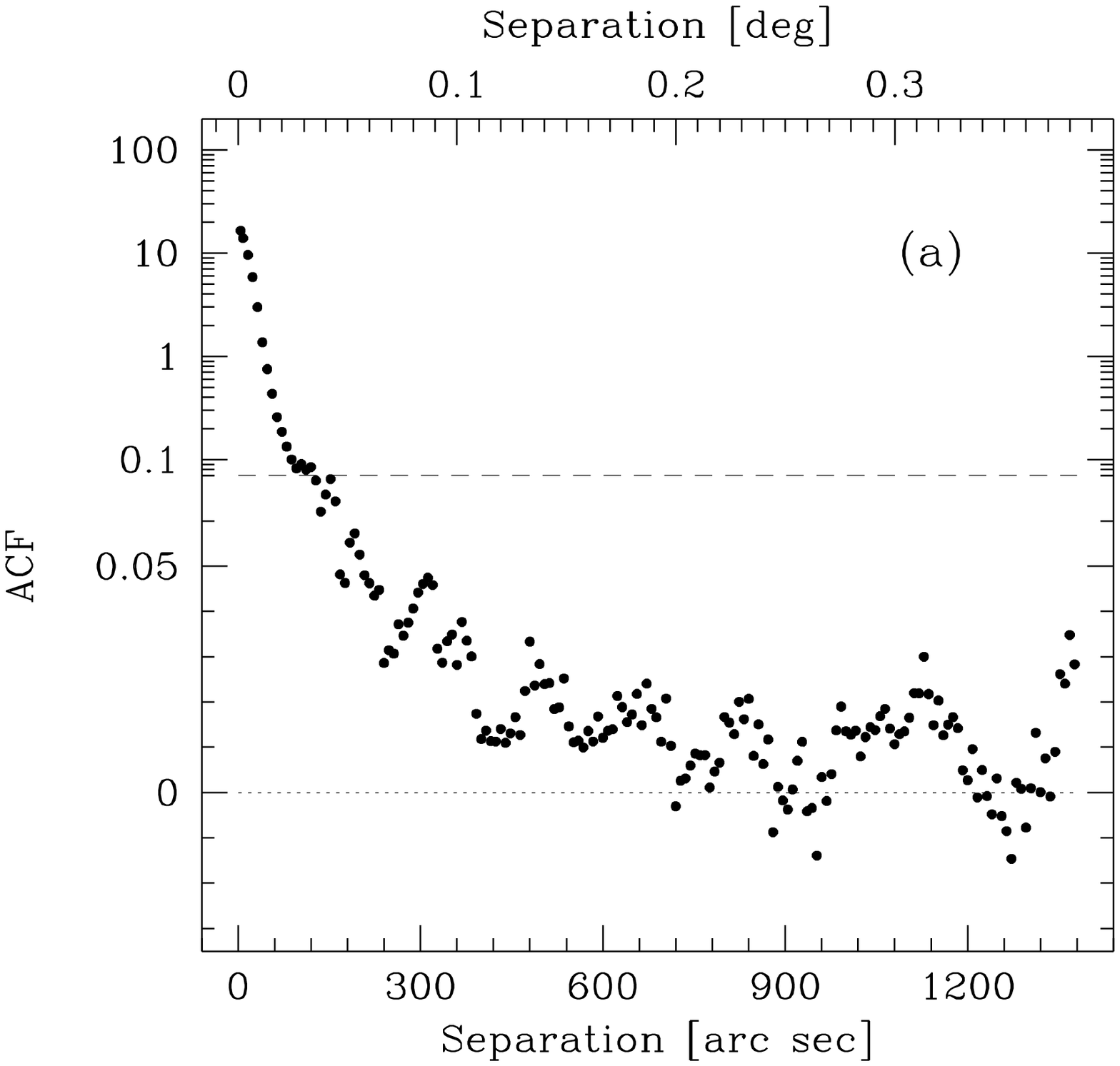}{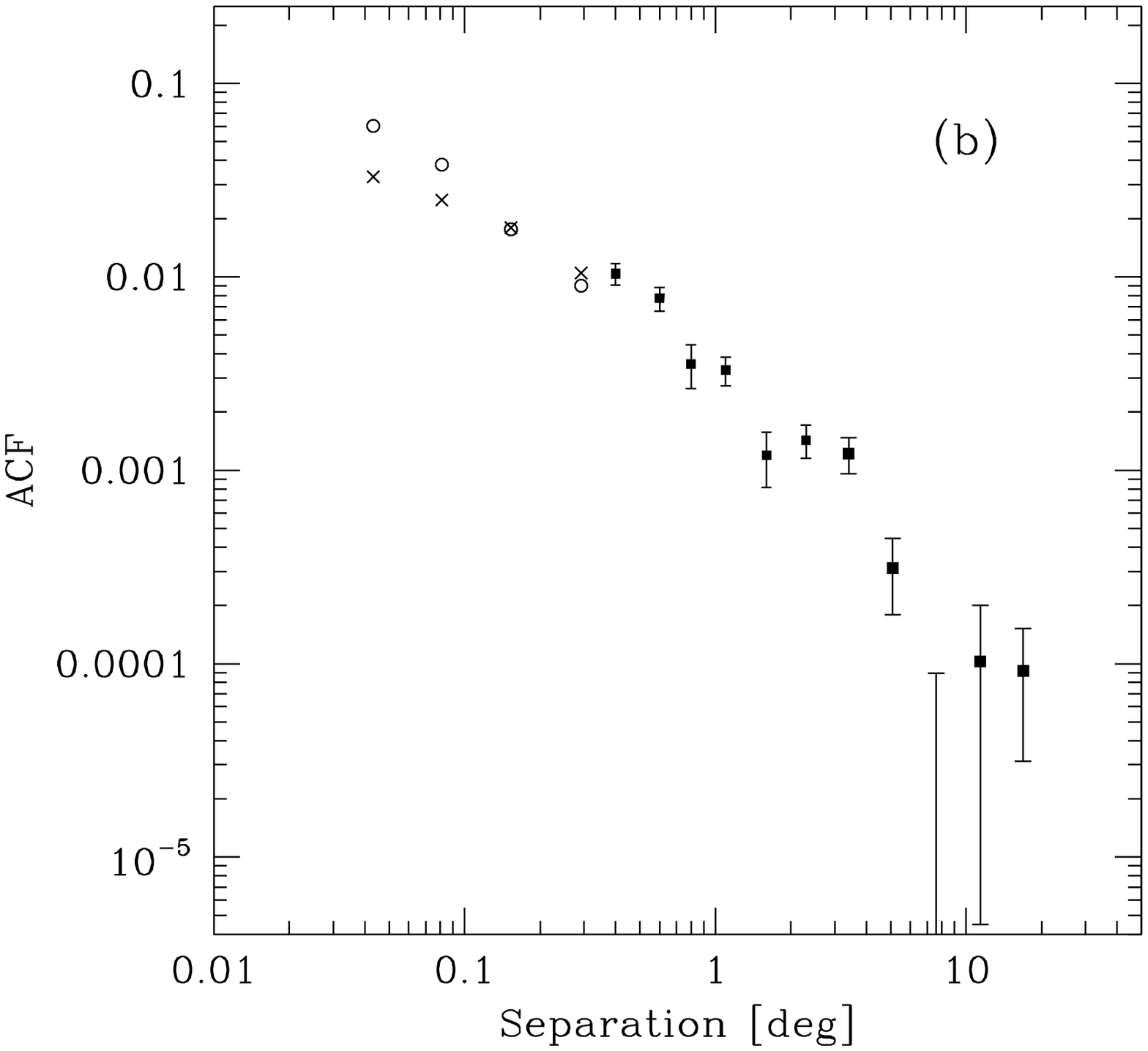}
\caption{a) The autocorrelation function of the soft XRB at small
separations based on {\it ROSAT} pointings (note logarithmic and linear
scales of the ordinate axis). \hfill \\
b) Estimates of the ACF based on the
RASS and pointings. Dots with the error bars show the RASS data; open
circles - all 141 pointings, crosses - only fields without bright clusters
(133 pointings)}
\end{figure}

\acknowledgments
This paper was supported by the Polish KBN grant 2~P03D~002~14.

\end{document}